\documentclass[runningheads]{llncs}
\usepackage{graphicx}
\usepackage{amsmath}
\usepackage{todonotes}
\usepackage{multirow}
\usepackage{float}
\begin{document}
\title{Memory-efficient Learning for High-Dimensional MRI Reconstruction}
\author{Ke Wang\inst{1} \and
Michael Kellman\inst{2} \and
Christopher M. Sandino\inst{3} \and Kevin Zhang\inst{1} \and Shreyas S. Vasanawala \inst{4}\and Jonathan I. Tamir \inst{5} \and Stella X. Yu\inst{6} \and Michael Lustig\inst{1}}
% %
\authorrunning{K. Wang et al.}
\institute{Electrical Engineering and Computer Sciences, UC Berkeley, Berkeley CA, USA \email{mikilustig@berkeley.edu} \and Pharmaceutical Chemistry, UCSF, San Francisco CA, USA \and Electrical Engineering, Stanford University, Stanford CA, USA \and Radiology, Stanford University, Stanford CA, USA  \and Electrical and Computer Engineering, UT Austin, Austin TX, USA \and International Computer Science Institute, UC Berkeley, Berkeley CA, USA}
% \email{mikilustig@berkeley.edu}\\
\maketitle              % typeset the header of the contribution
\begin{abstract}
Deep learning (DL) based unrolled reconstructions have shown state-of-the-art performance for under-sampled magnetic resonance imaging (MRI). Similar to compressed sensing, DL can leverage high-dimensional data (e.g. 3D, 2D+time, 3D+time) to further improve performance. However, network size and depth are currently limited by the GPU memory required for backpropagation. Here we use a memory-efficient learning (MEL) framework which favorably trades off 
storage with a manageable increase in computation during training. Using MEL with multi-dimensional data, we demonstrate improved image reconstruction performance for in-vivo 3D MRI and 2D+time cardiac cine MRI. MEL uses far less GPU memory while marginally increasing the training time, which enables new applications of DL to high-dimensional MRI.

\keywords{Magnetic Resonance Imaging (MRI)  \and Unrolled reconstruction \and Memory-efficient learning.}
\end{abstract}
%
%
%
% \newpage
\section{Introduction}
\begin{figure}[!ht]
\centering
\includegraphics[width=12cm]{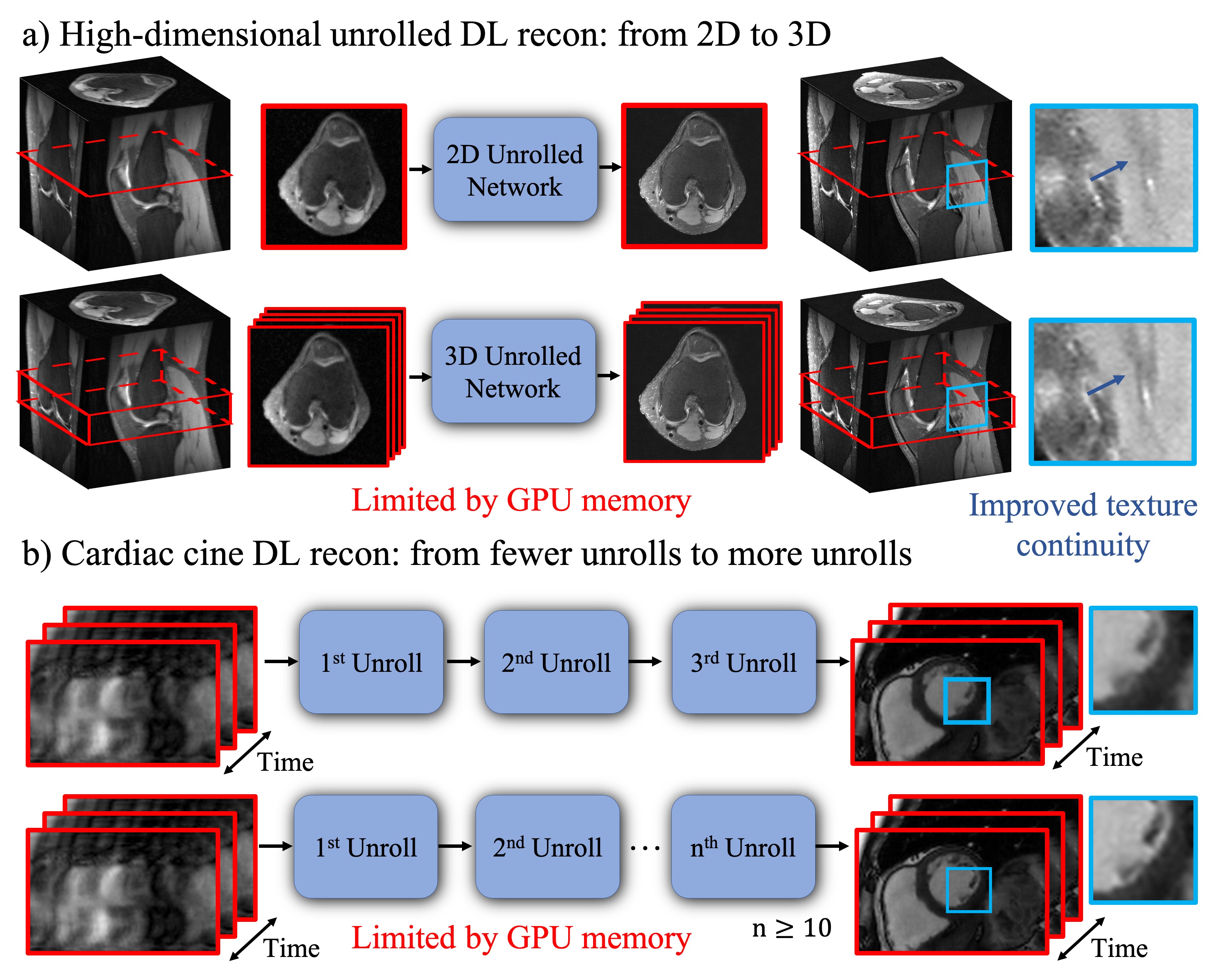}
% \caption{GPU memory limitations for high-dimensional unrolled DL recons: a) Conventionally, in order to reconstruct a 3D volume, each slice across the readout dimension passes through a 2D unrolled network and then is reformatted back into a 3D volume. In contrast, 3D unrolled networks require more memory but take 3D slabs during the training, which leverages more data redundancy. b) Currently, cardiac cine DL recons are often performed with a small number of unrolls due to memory limitations. More unrolls are able to better learn the finer spatial and temporal textures.} \label{fig1}
\caption{GPU memory limitations for high-dimensional unrolled DL recons: a) Compared to a 2D unrolled network, the 3D unrolled network uses a 3D slab during training to leverage more redundancy, but is limited by GPU memory. b) Cardiac cine DL recons are often performed with a small number of unrolls due to memory limitations.} \label{fig:overview}
\end{figure}

Deep learning-based unrolled reconstructions (Unrolled DL recons)\cite{diamond2017unrolled,schlemper2017deep,aggarwal2018modl,hammernik2018learning,tamir2019unsupervised,kustner2020cinenet} have shown great success at under-sampled MRI reconstruction, well beyond the capabilities of parallel imaging and compressed sensing (PICS)\cite{lustig2007sparse,griswold2002generalized,pruessmann1999sense}. These methods are often formulated by unrolling the iterations of an image reconstruction optimization\cite{hammernik2018learning,aggarwal2018modl,tamir2019unsupervised} and use a training set to learn an implicit regularization term represented by a deep neural network. It has been shown that increasing the number of unrolls improves upon finer spatial and temporal textures in the reconstruction\cite{hammernik2018learning,aggarwal2018modl,sandino2021accelerating}. Similar to compressed sensing and other low-dimensional representations, DL recons can take advantage of additional structure in very high-dimensional data (e.g. 3D, 2D+time, 3D+time) to further improve image quality. However, these large-scale DL recons are currently limited by GPU memory required for gradient-based optimization using backpropagation. Therefore, most Unrolled DL recons focus on 2D applications or are limited to a small number of unrolls. In this work, we use our recently proposed memory-efficient learning (MEL) framework\cite{kellman2020memory,zhang2020meld} to reduce the memory needed for backpropagation, which enables the training of Unrolled DL recons for 1) larger-scale 3D MRI; and 2) 2D+time cardiac cine MRI with a large number of unrolls (Figure \ref{fig:overview}). We evaluate the spatio-temporal complexity of our proposed method on the Model-based Deep Learning (MoDL) architecture \cite{aggarwal2018modl} and train these high-dimensional DL recons on a single 12GB GPU. Our training uses far less memory while only marginally increasing the computation time. To demonstrate the advantages of high-dimensional reconstructions to image quality, we performed experiments on both retrospectively and prospectively under-sampled data for 3D MRI and cardiac cine MRI. Our in-vivo experiments indicate that by exploiting high-dimensional data redundancy, we can achieve better quantitative metrics and improved image quality with sharper edges for both 3D MRI and cardiac cine MRI.

\section{Methods}
\subsection{Memory-efficient learning}
\begin{figure}[!ht]
\centering
\includegraphics[width=12.2cm]{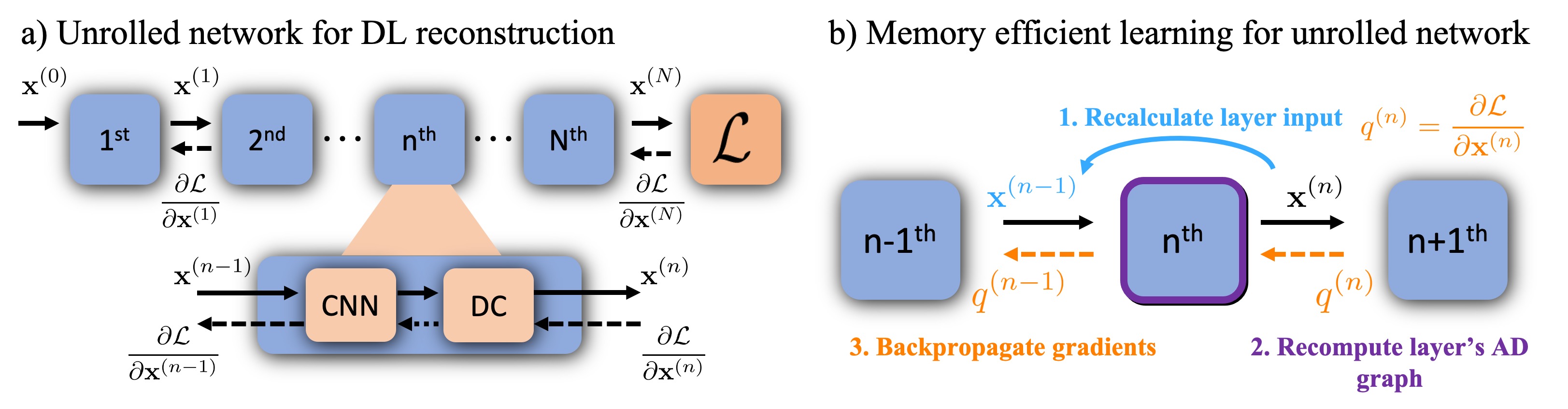}
\caption{a) In conventional DL recon training, gradients of all layers are evaluated as a single computational graph, requiring signifcant GPU memory. b) In MEL, we sequentially evaluate each layer by: i) Recalculate the layer’s input $\mathbf{x}^{(n-1)}$, from the known output $\mathbf{x}^{(n)}$. ii) Reform the AD graph for that layer. iii) Backpropagate gradients $q^{(n-1)}$ through the layer’s AD graph.} \label{fig:mel}
\end{figure}

As shown in Figure \ref{fig:mel} a), unrolled DL recons are often formulated by unrolling the iterations of an image reconstruction optimization\cite{hammernik2018learning,aggarwal2018modl}. Each unroll consists of two submodules: CNN based regularization layer and data consistency (DC) layer. In conventional backpropagation, the gradient must be computed for the entire computational graph, and intermediate variables from all $N$ unrolls need to be stored at a significant memory cost. By leveraging MEL, we can process the full graph as a series of smaller sequential graphs. As shown in Figure \ref{fig:mel} b), first, we forward propagate the network to get the output $\mathbf{x}^{(N)}$ without computing the gradients. Then, we rely on the invertibility of each layer (required) to recompute each smaller auto-differentiation (AD) graph from the network’s output in reverse order. MEL only requires a single layer to be stored in memory at a time, which reduces the required memory by a factor of $N$. Notably, the required additional computation to invert each layer only marginally increases the backpropagation runtime.

\subsection{Memory-efficient learning for MoDL}

Here, we use a widely used Unrolled DL Recon framework: MoDL\cite{aggarwal2018modl}. We formulate the reconstruction of $\mathbf{\hat{x}}$ as an optimization problem and solve it as below:
\begin{equation}
    \mathbf{\hat{x}}=\arg\min_\mathbf{x}\|\mathbf{Ax-y}\|^2_2+\mu\|\mathbf{x}-R_w(\mathbf{x})\|^2_2,
\end{equation}
where $\mathbf{A}$ is the system encoding matrix, $\mathbf{y}$ denotes the k-space measurements and $R_w$ is a learned CNN-based denoiser. For multi-channel MRI reconstruction, $\mathbf{A}$ can be formulated as $\mathbf{A} = \mathbf{PFS}$, where $\mathbf{S}$ represent the multi-channel sensitivity maps, $\mathbf{F}$ denotes Fourier Transform and $\mathbf{P}$ is the undersampling mask used for selecting the acquired data. MoDL solves the minimization problem by an alternating procedure:
\begin{equation}
    \mathbf{z}_n=R_w(\mathbf{x}_n)
\end{equation}

\begin{equation}
\begin{split}
    \mathbf{x}_{n+1}&=\arg\min_\mathbf{x}\|\mathbf{Ax-y}\|^2_2+\mu\|\mathbf{x}-\mathbf{z}_n\|^2_2,\\
    &= (\mathbf{A^HA+\mu\mathbf{I}})^{-1}(\mathbf{A^Hy}+\mu\mathbf{z}_n)
\end{split}
\end{equation}
which represents the CNN-based regularization layer and DC layer respectively. In this formulation, the DC layer is solved using Conjugate Gradient (CG)\cite{shewchuk1994introduction}, which is unrolled for a finite number of iterations. For all the experiments, we used an invertible residual convolutional neural network (RCNN) introduced in \cite{gomez2017reversible,ovadia2019can,he2016deep}, whose architecture is composed of a 5-layer CNN with 64 channels per layer. Detailed network architecture is shown in Figure \ref{fig:figS1}. The residual CNN is inverted using the fixed-point algorithm as described in \cite{kellman2020memory}, while the DC layer is inverted through:
\begin{equation}
    \mathbf{z}_n=\frac{1}{\mu}((\mathbf{A^HA}+\mu\mathbf{I})\mathbf{x}_{n+1}-\mathbf{A^Hy}).
\end{equation}

\subsection{Training and evaluation of memory-efficient learning}
With IRB approval and informed consent/assent, we trained and evaluated MEL on both retrospective and prospective 3D knee and 2D+time cardiac cine MRI. We conducted 3D MoDL experiments with and without MEL on 20 fully-sampled 3D knee datasets (320 slices each) from mridata.org\cite{sawyer2013creation}. 16 cases were used for training, 2 cases were used for validation and other 2 for testing. Around 5000 3D slabs with size 21$\times$256$\times$320 were used for training the reconstruction networks. All data were acquired on a 3T GE Discovery MR 750 with an 8-channel HD knee coil. An 8x Poisson Disk sampling pattern was used to retrospectively undersample the fully sampled k-space. Scan parameters included a matrix size of 320$\times$256$\times$320, and TE/TR of 25ms/1550ms. In order to further demonstrate the feasibility of our 3D reconstruction with MEL on realistic prospectively undersampled scans, we reconstructed 8$\times$ prospectively undersampled 3D FSE knee scans (available at mridata.org) with the model trained on retrospectively undersampled knee data. Scanning parameters includes: Volume size: 320$\times$288$\times$236, TR/TE = 1400/20.46ms, Flip Angle: 90$^{\circ}$, FOV: 160 mm$\times$160 mm$\times$ 141.6 mm.

For the cardiac cine MRI, fully-sampled bSSFP cardiac cine datasets were acquired from 15 volunteers at different cardiac views and slice locations on 1.5T and 3.0T GE scanners using a 32-channel cardiac coil. All data were coil compressed\cite{zhang2013coil} to 8 virtual coils. Twelve of the datasets (around 190 slices) were used for training, 2 for validation, and one for testing. k-Space data were retrospectively under-sampled using a variable-density k-t sampling pattern to simulate 14-fold acceleration with 25\% partial echo. We also conducted experiments on a prospectively under-sampled scan (R=12) which was acquired from a pediatric patient within a single breath-hold on a 1.5T scanner.

We compared the spatio-temporal complexity (GPU memory, training time) with and without MEL. In order to show the benefits of high-dimensional DL recons, we compared the reconstruction results of PICS, 2D and 3D MoDL with MEL for 3D MRI, and 2D+time MoDL with 4 unrolls and 10 unrolls for cardiac cine MRI. For both 2D MoDL and 3D MoDL with MEL, we used 5 unrolls, 10 CG steps and Residual CNN as the regularization layer. A baseline PICS reconstruction was performed using BART\cite{uecker2015berkeley}. Sensitivity maps were computed using BART\cite{uecker2015berkeley} and SigPy\cite{ong2019sigpy}. Common image quality metrics such as Peak Signal to Noise Ratio (pSNR), Structual Similarity (SSIM) \cite{isola2017image} and Fréchet Inception Distance (FID)\cite{heusel2017gans} were reported. FID is a widely used measure of perceptual similarity between two sets of images. All the experiments were implemented in Pytorch \cite{NEURIPS2019_9015} and used Nvidia Titan XP (12GB) and Titan V CEO (32GB) GPUs. Networks were trained end-to-end using a per-pixel $l_1$ loss and optimized using Adam \cite{kingma2014adam} with a learning rate of $1\times10^{-4}$.

\section{Results}

\begin{figure}[!ht]
\centering
\includegraphics[width=10.2cm]{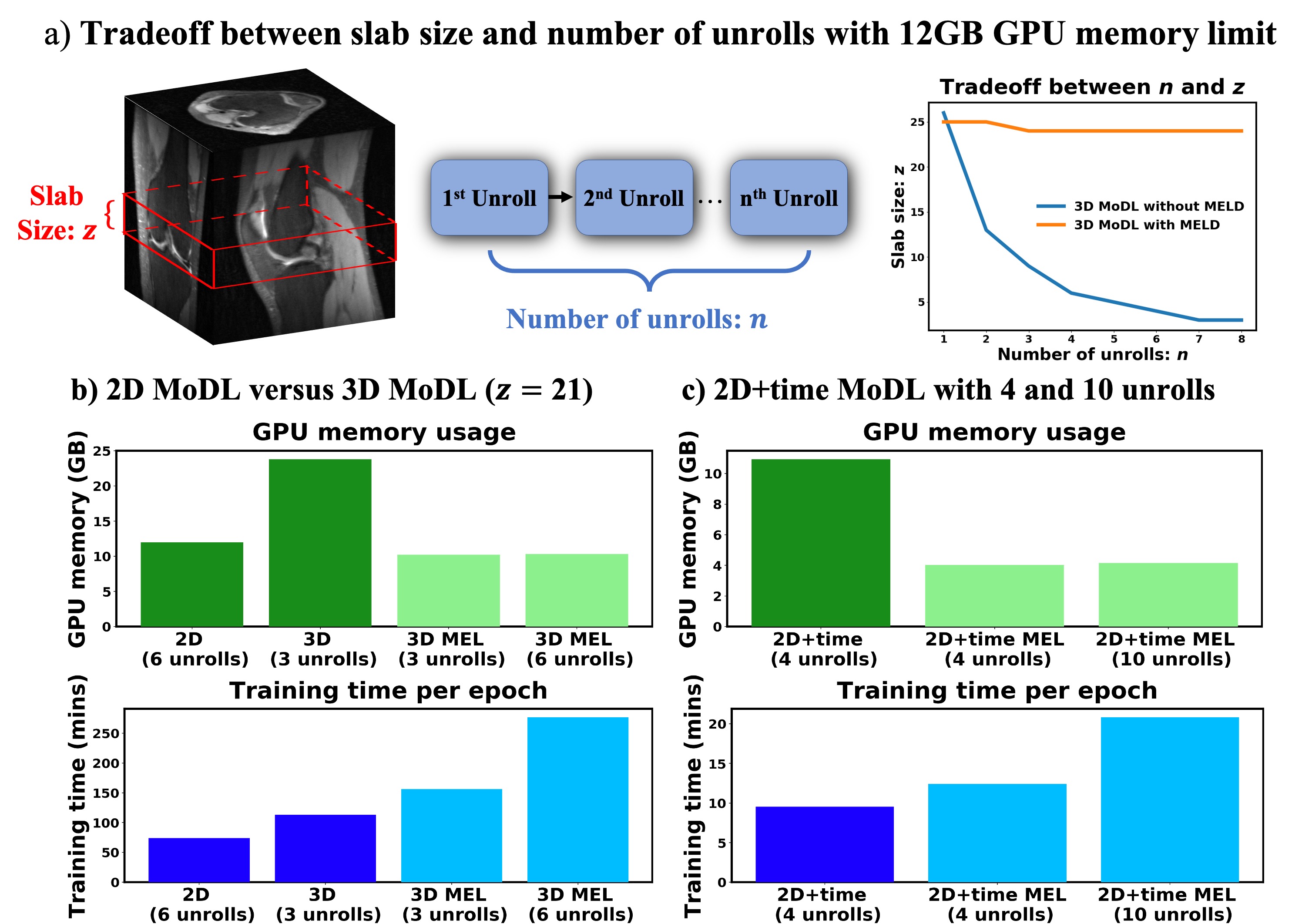}
\caption{Spatio-temporal complexity of MoDL with and without MEL. a) Tradeoff between 3D slab size $z$ and a number of unrolls $n$ with a 12GB GPU memory limitation. b) and c) show the memory and time comparisons for MoDL with and without MEL.
} \label{fig:fig3}
\end{figure}

We first evaluate the spatio-temporal complexity of MoDL with and without MEL (Figure \ref{fig:fig3}). Without MEL, for a 12GB GPU memory limit, the maximum slab size decreases rapidly as the number of unrolls increases, which limits the performance of a 3D reconstruction. In contrast, using MEL, the maximum slab size is roughly constant. Figure \ref{fig:fig3} b) and c) show the comparisons from two different perspectives: 1)GPU memory usage; 2)Training time per epoch. Results indicate that for both 3D and 2D+time MoDL,  MEL uses significantly less GPU memory than conventional backpropagation while marginally increasing training time. Notably, both MoDL with and without MEL have the same inference time.

\begin{figure}[!ht]
\centering
\includegraphics[width=12cm]{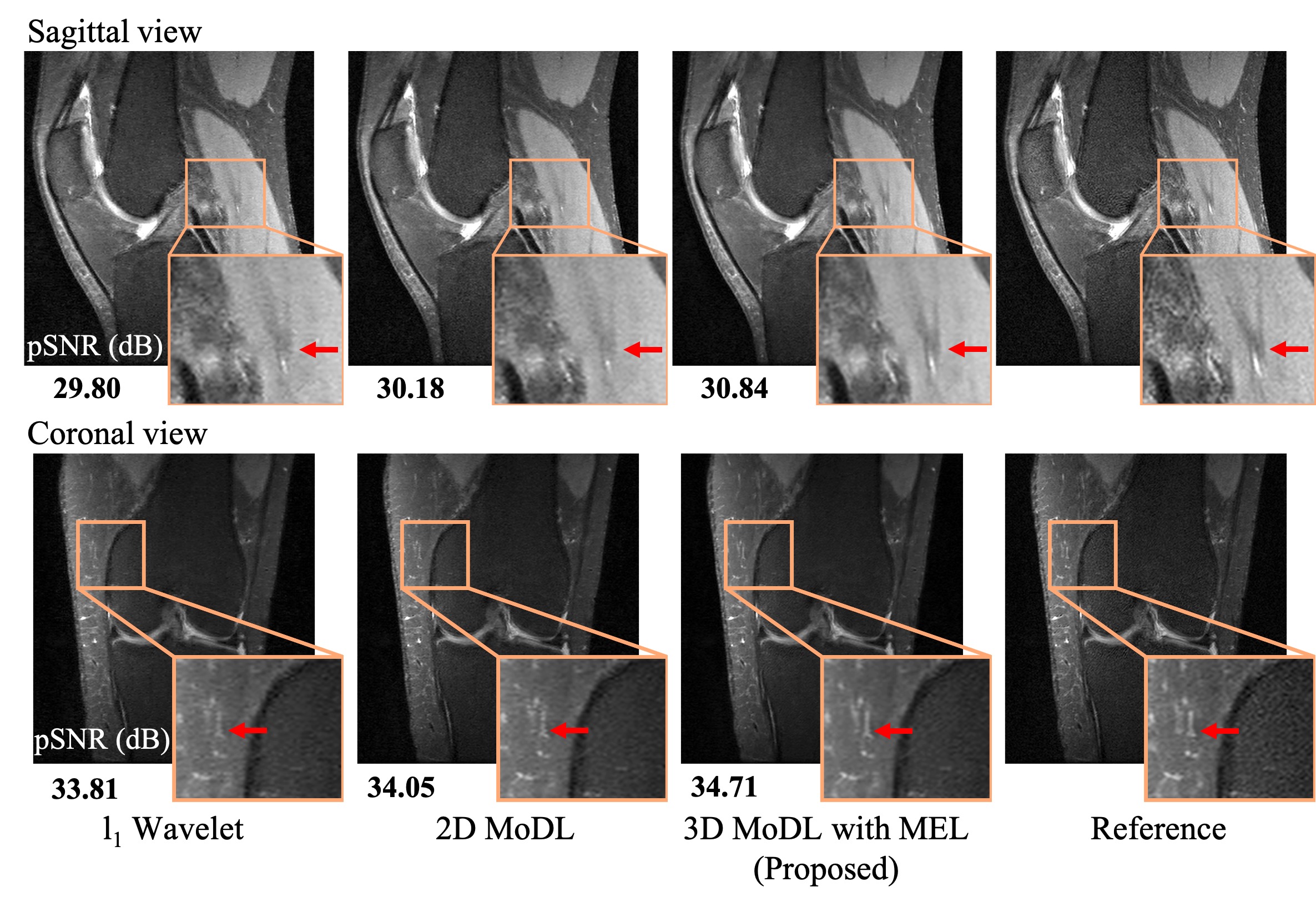}
\caption{A representative comparison of different methods (PICS, 2D MoDL, 3D MoDL with MEL) on 3D knee reconstruction (Sagittal view and Coronal view are shown). pSNRs are shown under each reconstructed image.} 
\label{fig:fig4}
\end{figure}

\begin{figure}[ht]
\centering
\includegraphics[width=12cm]{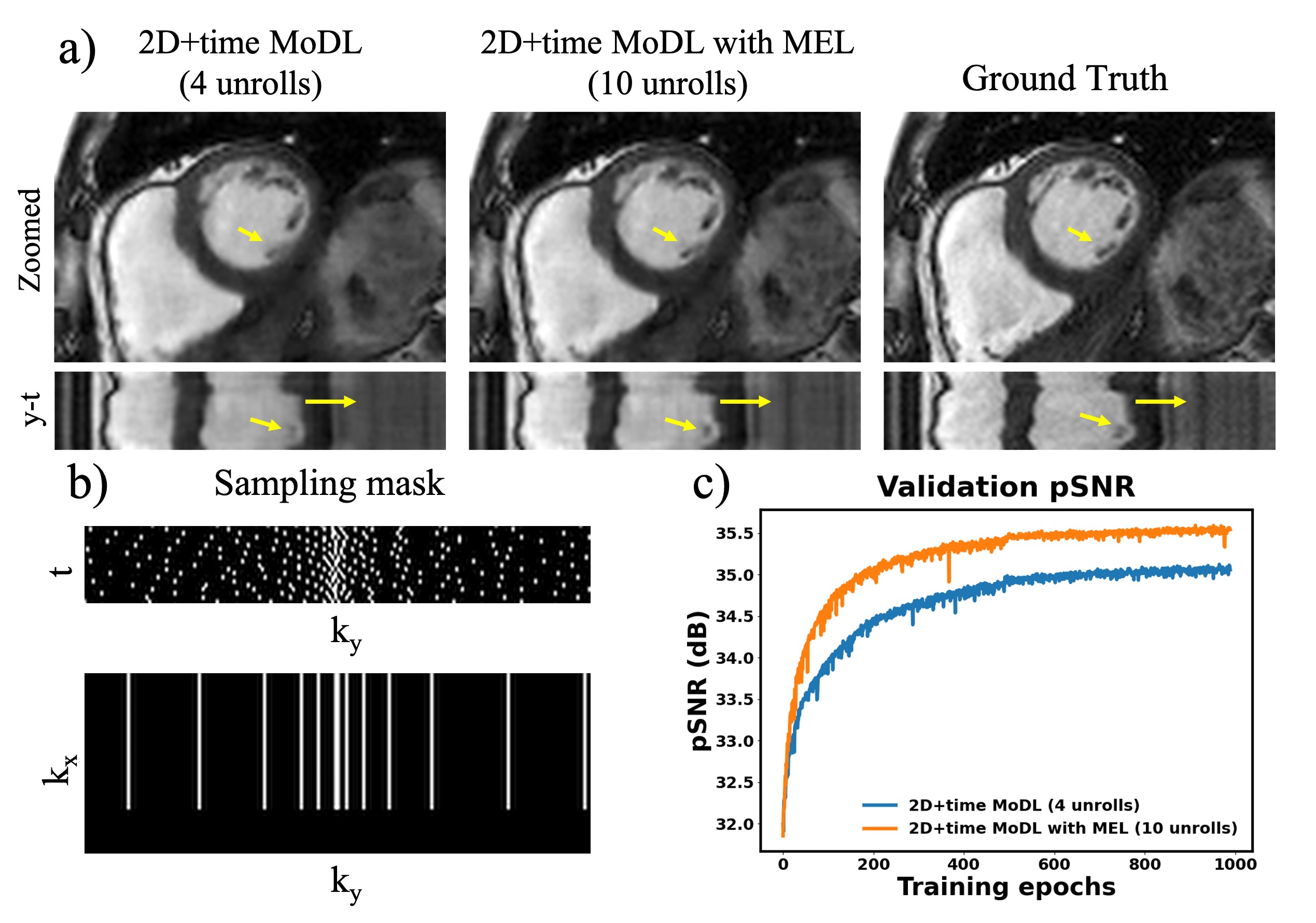}
\caption{a) Short-axis view cardiac cine reconstruction of a healthy volunteer on a 1.5T scanner. k-Space data was retrospectively undersampled to simulate 14-fold acceleration with 25\% partial echo (shown in b) and reconstructed by: 2D+time MoDL with 4 unrolls, 2D+time MoDL with MEL and 10 unrolls. c) Validation pSNR of MoDL with 4 unrolls and MoDL with 10 unrolls.
} 
\label{fig:fig5}
\end{figure}

Figure \ref{fig:fig4} shows a comparison of different methods for 3D reconstruction. Instead of learning from only 2D axial view slices (Figure \ref{fig:overview} a), 3D MoDL with MEL captures the image features from all three dimensions. Zoomed-in details indicate that 3D MoDL with MEL is able to provide more faithful contrast with more continuous and realistic textures as well as higher pSNR over other methods. Figure \ref{fig:fig5} demonstrates that MEL enables the training of 2D+time MoDL with a large number of unrolls (10 unrolls), which outperforms MoDL with 4 unrolls with respect to image quality and y-t motion profile. With MEL, MoDL with 10 unrolls resolves the papillary muscles (yellow arrows) better than MoDL with 4 unrolls. Also, the y-t profile of MoDL with 10 unrolls depicts motion in a more natural way while MoDL with 4 unrolls suffers from blurring. Meanwhile, using 10 unrolls over 4 unrolls yields an improvement of 0.6dB in validation pSNR.

Table \ref{tab:my_label} shows the quantitative metric comparisons (pSNR, SSIM and FID) between different methods on both 3D MRI and cardiac cine MRI reconstructions. 
% For 3D MRI, metrics are computed on all sagittal slices in the testing set with respect to fully-sampled images. For 2D Cardiac Cine MRI, metrics are computed on slices from 3 different orientations and averaged across time dimension. 
The results indicate that both 3D MoDL with MEL and 2D+time MoDL with MEL outperform other methods with respect to pSNR, SSIM and FID.

\begin{table}[!ht]
    \centering
        \begin{center}
        \begin{tabular}{ c|c|c|c } 
        \hline
        metric & method & 3D MRI & 2D cardiac cine MRI \\
        \hline
        \multirow{5}{4em}{pSNR (dB)} & PICS & 31.01$\pm$1.97 & 24.69$\pm$2.74 \\ 
        & 2D MoDL & 31.44$\pm$2.07 & - \\ 
        & 3D MoDL with MEL & \textbf{32.11$\pm$2.05} & - \\ 
        & 2D+time MoDL: 4 unrolls & - & 26.87$\pm$2.98 \\ 
        & 2D+time MoDL with MEL: 10 unrolls & - & \textbf{27.42$\pm$3.21}\\
        \hline
        \multirow{5}{4em}{SSIM} & PICS & 0.816$\pm$0.046 & 0.824$\pm$0.071 \\ 
        & 2D MoDL & 0.821$\pm$0.044&- \\ 
        & 3D MoDL with MEL & \textbf{0.830$\pm$0.038}&- \\ 
        & 2D+time MoDL: 4 unrolls & - & 0.870$\pm$0.042 \\ 
        & 2D+time MoDL with MEL: 10 unrolls & - & \textbf{0.888$\pm$0.042}\\
        \hline
        \multirow{5}{4em}{FID} & PICS & 46.71& 39.40 \\ 
        & 2D MoDL & 43.58&- \\ 
        & 3D MoDL with MEL & \textbf{41.48}&- \\ 
        & 2D+time MoDL: 4 unrolls & - & 36.93 \\ 
        & 2D+time MoDL with MEL: 10 unrolls & - & \textbf{31.64}\\
        \hline
        \end{tabular}
        \end{center}
    
    \caption{Quantitative metrics (pSNR, SSIM and FID) of different methods on 3D MRI and cardiac cine MRI reconstructions (mean $\pm$ standard deviation of pSNR and SSIM).}
    \label{tab:my_label}
\end{table}

\begin{figure}[!ht]
\centering
\includegraphics[width=12cm]{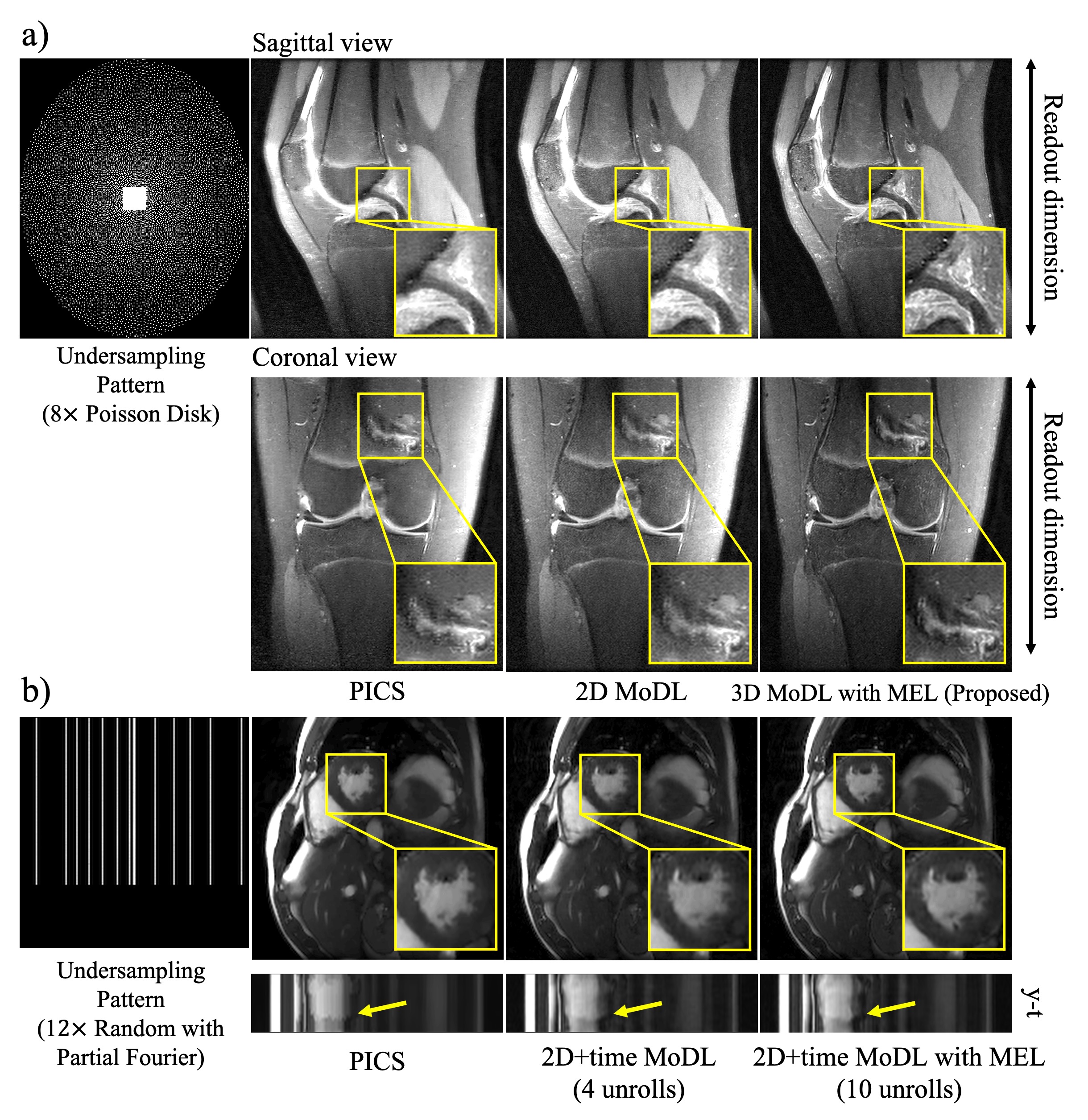}
\caption{a) Representative reconstruction results on a prospectively undersampled 3D FSE knee scan using different methods (PICS, 2D MoDL and 3D MoDL with MEL). b) Representative reconstruction results on a prospectively undersampled cardiac cine dataset. y-t motion profiles are shown along with the reconstructed images.
} 
\label{fig:fig6}
\end{figure}

Figure \ref{fig:fig6} a) and Figure \ref{fig:figS2} show the reconstruction results on two representative prospectively undersampled 3D FSE knee scan. Note that in this scenario, there is no fully-sampled groud truth. Despite there exists some difference between the training and testing (e.g., matrix size, scanning parameters), 3D MoDL with MEL is still able to resolve more detailed texture and sharper edges over traditional PICS and learning-based 2D MoDL. Figure \ref{fig:fig6} b) and Video \ref{fig:vidS1} shows the reconstruction on a representative prospective undersampled cardiac cine scan. We can clearly see that enabled by MEL, 2D+time MoDL with 10 unrolls can better depicts the finer details as well as more natural motion profile.

\section{Conclusions}
In this work, we show that MEL enables learning for high-dimensional MR reconstructions on a single 12GB GPU, which is not possible with standard backpropagation methods. We demonstrate MEL on two representative large-scale MR reconstruction problems: 3D volumetric MRI, 2D cardiac cine MRI with a relatively large number of unrolls. By leveraging the high-dimensional image redundancy and a large number of unrolls, we were able to get improved quantitative metrics and reconstruct finer details, sharper edges, and more continuous textures with higher overall image quality for both 3D and 2D cardiac cine MRI. Furthermore, 3D MoDL reconstruction results from prospectively undersampled k-space show that the proposed method is robust to the scanning parameters and could be potentially deployed in clinical systems. Overall, MEL brings a practical tool for training the large-scale high-dimensional MRI reconstructions with much less GPU memory and is able to achieve improved reconstructed image quality.

\section{Acknowledgements}
The authors would like to thank Dr. Gopal Nataraj for his helpful discusses and paper editing. We also acknowledge support from NIH R01EB009690, NIH R01HL136965, NIH R01EB026136 and GE Healthcare.
% \begin{theorem}
% This is a sample theorem. The run-in heading is set in bold, while
% the following text appears in italics. Definitions, lemmas,
% propositions, and corollaries are styled the same way.
% \end{theorem}
% %
% % the environments 'definition', 'lemma', 'proposition', 'corollary',
% % 'remark', and 'example' are defined in the LLNCS documentclass as well.
% %
% \begin{proof}
% Proofs, examples, and remarks have the initial word in italics,
% while the following text appears in normal font.
% \end{proof}
% For citations of references, we prefer the use of square brackets
% and consecutive numbers. Citations using labels or the author/year
% convention are also acceptable. The following bibliography provides
% a sample reference list with entries for journal
% articles~\cite{ref_article1}, an LNCS chapter~\cite{ref_lncs1}, a
% book~\cite{ref_book1}, proceedings without editors~\cite{ref_proc1},
% and a homepage~\cite{ref_url1}. Multiple citations are grouped
% \cite{ref_article1,ref_lncs1,ref_book1},
% \cite{ref_article1,ref_book1,ref_proc1,ref_url1}.
\clearpage
\newpage
\bibliographystyle{splncs04}
\bibliography{ref}

\begin{thebibliography}{10}
\providecommand{\url}[1]{\texttt{#1}}
\providecommand{\urlprefix}{URL }
\providecommand{\doi}[1]{https://doi.org/#1}

\bibitem{aggarwal2018modl}
Aggarwal, H.K., Mani, M.P., Jacob, M.: Modl: Model-based deep learning
  architecture for inverse problems. IEEE transactions on medical imaging
  \textbf{38}(2),  394--405 (2018)

\bibitem{diamond2017unrolled}
Diamond, S., Sitzmann, V., Heide, F., Wetzstein, G.: Unrolled optimization with
  deep priors. arXiv preprint arXiv:1705.08041  (2017)

\bibitem{gomez2017reversible}
Gomez, A.N., Ren, M., Urtasun, R., Grosse, R.B.: The reversible residual
  network: Backpropagation without storing activations. arXiv preprint
  arXiv:1707.04585  (2017)

\bibitem{griswold2002generalized}
Griswold, M.A., Jakob, P.M., Heidemann, R.M., Nittka, M., Jellus, V., Wang, J.,
  Kiefer, B., Haase, A.: Generalized autocalibrating partially parallel
  acquisitions (grappa). Magnetic Resonance in Medicine: An Official Journal of
  the International Society for Magnetic Resonance in Medicine  \textbf{47}(6),
   1202--1210 (2002)

\bibitem{hammernik2018learning}
Hammernik, K., Klatzer, T., Kobler, E., Recht, M.P., Sodickson, D.K., Pock, T.,
  Knoll, F.: Learning a variational network for reconstruction of accelerated
  mri data. Magnetic resonance in medicine  \textbf{79}(6),  3055--3071 (2018)

\bibitem{he2016deep}
He, K., Zhang, X., Ren, S., Sun, J.: Deep residual learning for image
  recognition. In: Proceedings of the IEEE conference on computer vision and
  pattern recognition. pp. 770--778 (2016)

\bibitem{heusel2017gans}
Heusel, M., Ramsauer, H., Unterthiner, T., Nessler, B., Hochreiter, S.: Gans
  trained by a two time-scale update rule converge to a local nash equilibrium.
  arXiv preprint arXiv:1706.08500  (2017)

\bibitem{isola2017image}
Isola, P., Zhu, J.Y., Zhou, T., Efros, A.A.: Image-to-image translation with
  conditional adversarial networks. In: Proceedings of the IEEE conference on
  computer vision and pattern recognition. pp. 1125--1134 (2017)

\bibitem{kellman2020memory}
Kellman, M., Zhang, K., Markley, E., Tamir, J., Bostan, E., Lustig, M., Waller,
  L.: Memory-efficient learning for large-scale computational imaging. IEEE
  Transactions on Computational Imaging  \textbf{6},  1403--1414 (2020)

\bibitem{kingma2014adam}
Kingma, D.P., Ba, J.: Adam: A method for stochastic optimization. arXiv
  preprint arXiv:1412.6980  (2014)

\bibitem{kustner2020cinenet}
K{\"u}stner, T., Fuin, N., Hammernik, K., Bustin, A., Qi, H., Hajhosseiny, R.,
  Masci, P.G., Neji, R., Rueckert, D., Botnar, R.M., et~al.: Cinenet: deep
  learning-based 3d cardiac cine mri reconstruction with multi-coil
  complex-valued 4d spatio-temporal convolutions. Scientific reports
  \textbf{10}(1),  1--13 (2020)

\bibitem{lustig2007sparse}
Lustig, M., Donoho, D., Pauly, J.M.: Sparse mri: The application of compressed
  sensing for rapid mr imaging. Magnetic Resonance in Medicine: An Official
  Journal of the International Society for Magnetic Resonance in Medicine
  \textbf{58}(6),  1182--1195 (2007)

\bibitem{ong2019sigpy}
Ong, F., Lustig, M.: Sigpy: a python package for high performance iterative
  reconstruction. In: Proc. ISMRM (2019)

\bibitem{ovadia2019can}
Ovadia, Y., Fertig, E., Ren, J., Nado, Z., Sculley, D., Nowozin, S., Dillon,
  J.V., Lakshminarayanan, B., Snoek, J.: Can you trust your model's
  uncertainty? evaluating predictive uncertainty under dataset shift. arXiv
  preprint arXiv:1906.02530  (2019)

\bibitem{NEURIPS2019_9015}
Paszke, A., Gross, S., Massa, F., Lerer, A., Bradbury, J., Chanan, G., Killeen,
  T., Lin, Z., Gimelshein, N., Antiga, L., Desmaison, A., Kopf, A., Yang, E.,
  DeVito, Z., Raison, M., Tejani, A., Chilamkurthy, S., Steiner, B., Fang, L.,
  Bai, J., Chintala, S.: Pytorch: An imperative style, high-performance deep
  learning library. In: Wallach, H., Larochelle, H., Beygelzimer, A.,
  d\textquotesingle Alch\'{e}-Buc, F., Fox, E., Garnett, R. (eds.) Advances in
  Neural Information Processing Systems 32, pp. 8024--8035. Curran Associates,
  Inc. (2019),
  \url{http://papers.neurips.cc/paper/9015-pytorch-an-imperative-style-high-performance-deep-learning-library.pdf}

\bibitem{pruessmann1999sense}
Pruessmann, K.P., Weiger, M., Scheidegger, M.B., Boesiger, P.: Sense:
  sensitivity encoding for fast mri. Magnetic Resonance in Medicine: An
  Official Journal of the International Society for Magnetic Resonance in
  Medicine  \textbf{42}(5),  952--962 (1999)

\bibitem{sandino2021accelerating}
Sandino, C.M., Lai, P., Vasanawala, S.S., Cheng, J.Y.: Accelerating cardiac
  cine mri using a deep learning-based espirit reconstruction. Magnetic
  Resonance in Medicine  \textbf{85}(1),  152--167 (2021)

\bibitem{sawyer2013creation}
Sawyer, A.M., Lustig, M., Alley, M., Uecker, P., Virtue, P., Lai, P.,
  Vasanawala, S.: Creation of fully sampled mr data repository for compressed
  sensing of the knee. Citeseer  (2013)

\bibitem{schlemper2017deep}
Schlemper, J., Caballero, J., Hajnal, J.V., Price, A.N., Rueckert, D.: A deep
  cascade of convolutional neural networks for dynamic mr image reconstruction.
  IEEE transactions on Medical Imaging  \textbf{37}(2),  491--503 (2017)

\bibitem{shewchuk1994introduction}
Shewchuk, J.R., et~al.: An introduction to the conjugate gradient method
  without the agonizing pain (1994)

\bibitem{tamir2019unsupervised}
Tamir, J.I., Yu, S.X., Lustig, M.: Unsupervised deep basis pursuit: Learning
  inverse problems without ground-truth data. arXiv preprint arXiv:1910.13110
  (2019)

\bibitem{uecker2015berkeley}
Uecker, M., Ong, F., Tamir, J.I., Bahri, D., Virtue, P., Cheng, J.Y., Zhang,
  T., Lustig, M.: Berkeley advanced reconstruction toolbox. In: Proc. Intl.
  Soc. Mag. Reson. Med. No.~2486 in 23 (2015)

\bibitem{zhang2020meld}
Zhang, K., Kellman, M., Tamir, J.I., Lustig, M., Waller, L.: Memory-efficient
  learning for unrolled 3d mri reconstructions. In: ISMRM Workshop on Data
  Sampling and Image Reconstruction (2020)

\bibitem{zhang2013coil}
Zhang, T., Pauly, J.M., Vasanawala, S.S., Lustig, M.: Coil compression for
  accelerated imaging with cartesian sampling. Magnetic resonance in medicine
  \textbf{69}(2),  571--582 (2013)

\end{thebibliography}

\newpage
\section{Supplementary Material}
\renewcommand{\thefigure}{S1}
\begin{figure}[!ht]
\centering
\includegraphics[width=12cm]{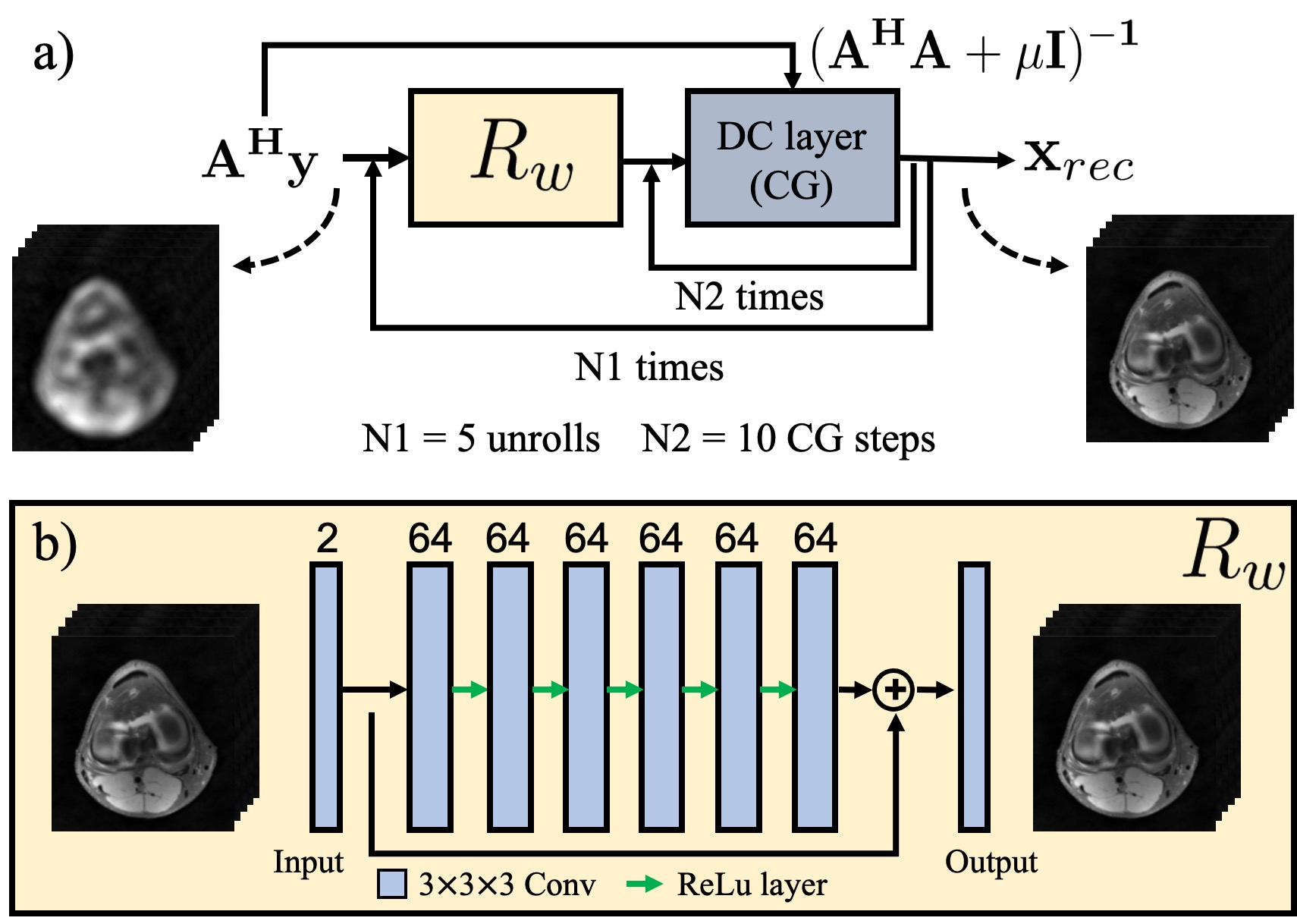}
\caption{Architectures of 3D MoDL framework and CNN based regularization layer. a) Zero-filled reconstruction is passed through N1 unrools consisting of a CNN based denoiser $R_w$ and N2 CG update steps. b) The CNN uses a Residual CNN architecture with convolutional blocks (3$\times$3$\times$3 kernel, 64 channel) and ReLU activations, which has been proved to be invertible. 
} 
\label{fig:figS1}
\end{figure}

\renewcommand{\figurename}{Video.}
\renewcommand{\thefigure}{S1}
\begin{figure}[!ht]
\centering
\includegraphics[width=12cm]{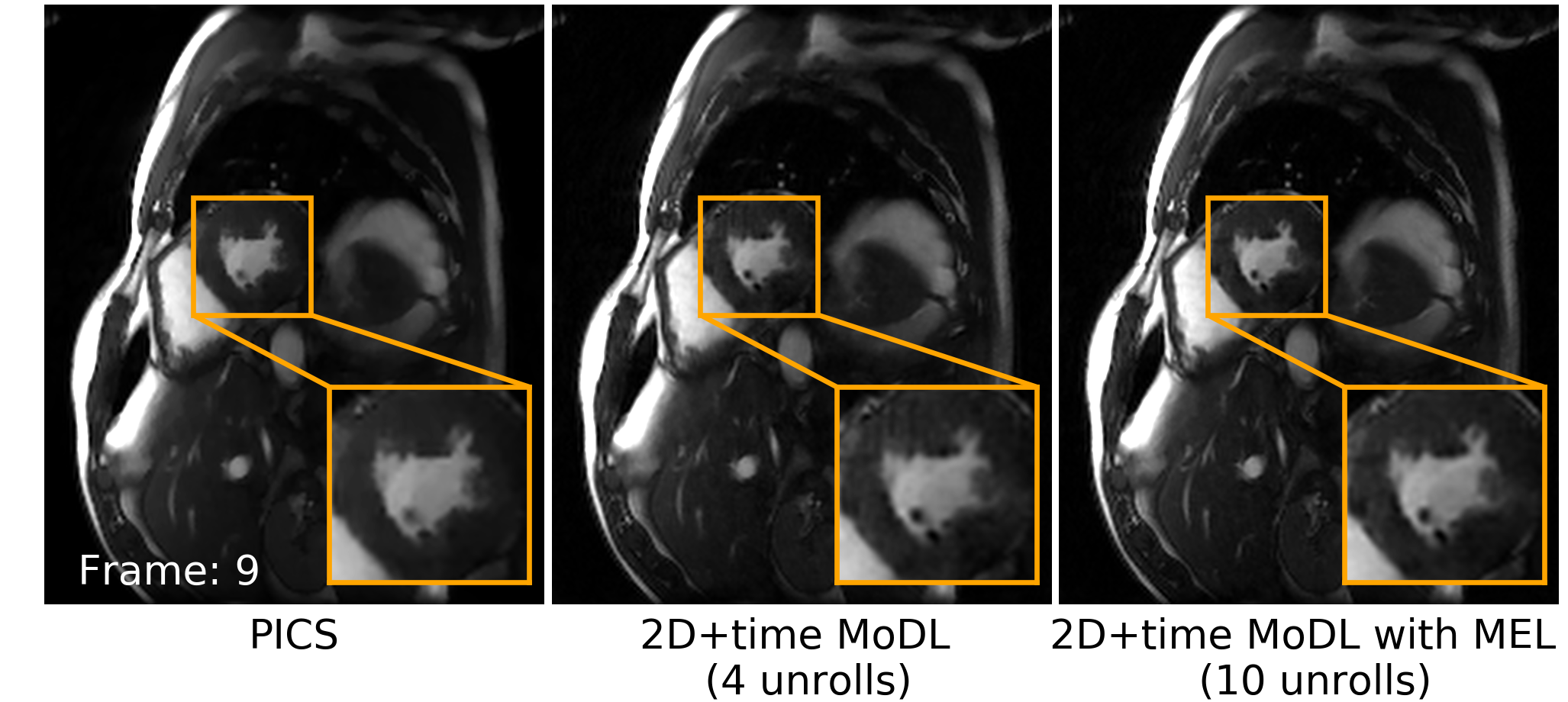}
\caption{A dynamic video for a representative reconstruction results on a prospectively undersampled cardiac cine scan (See attached MP4 file). 
} 
\label{fig:vidS1}
\end{figure}

\renewcommand{\figurename}{Fig.}
\renewcommand{\thefigure}{S2}
\begin{figure}[!ht]
\centering
\includegraphics[width=12cm]{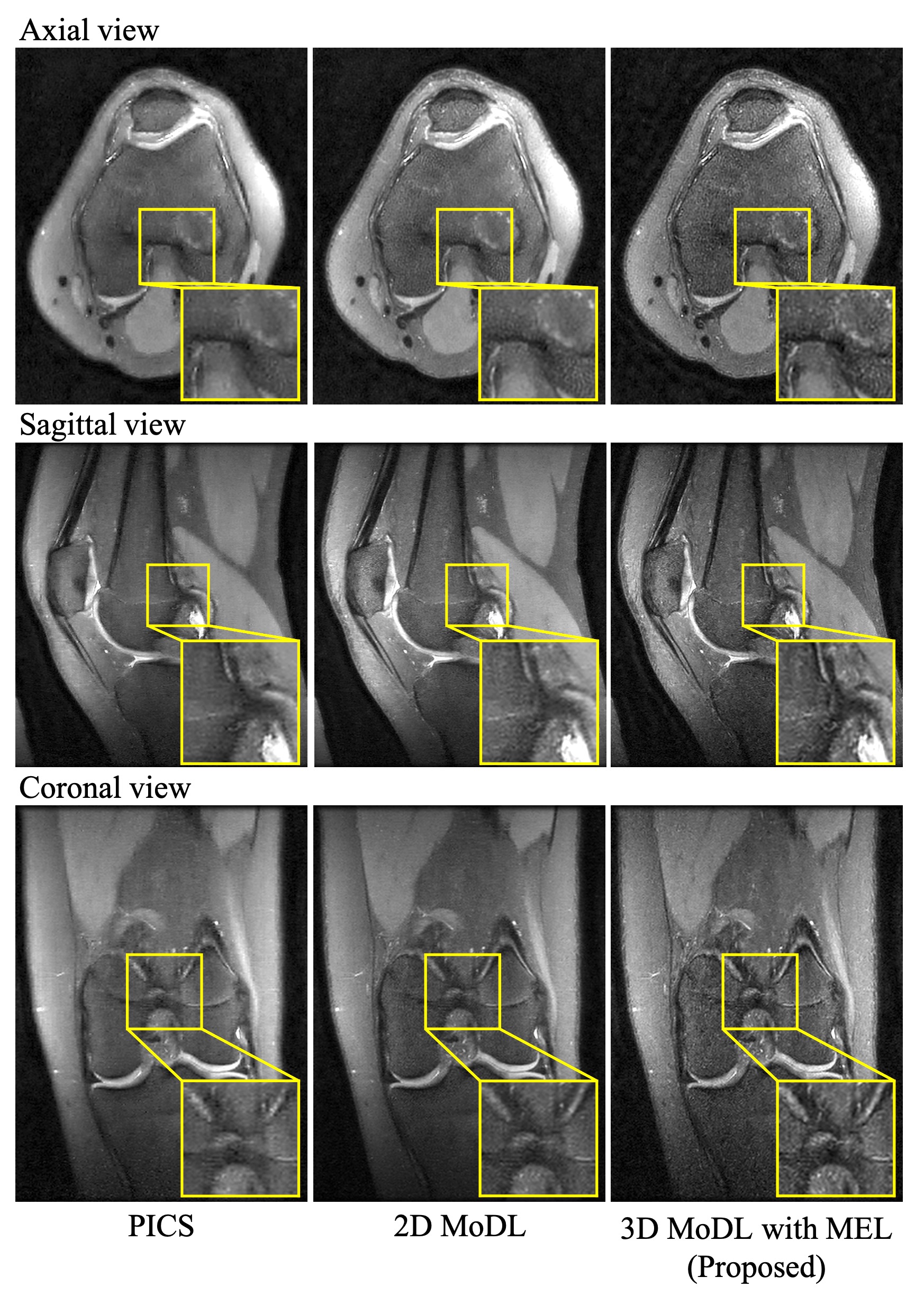}
\caption{A representative reconstruction results on prospectively undersampled 3D FSE knee scan with the model trained on retrospectively undersampled knee data using different methods (PICS, 2D MoDL and 3D MoDL with MEL). Representative slices from 3 different views (Axial, Sagittal, Coronal) are shown here. 
} 
\label{fig:figS2}
\end{figure}

\end{document}